\documentclass[a4paper,twocolumn,10pt,unpublished
]{quantumarticle}
\pdfoutput=1
\usepackage[utf8]{inputenc}
\usepackage[english]{babel}
\usepackage[T1]{fontenc}
\usepackage{amsmath}
\usepackage{amssymb}
\usepackage{hyperref}
\usepackage[numbers,sort&compress]{natbib}
\usepackage{braket}
\usepackage{graphicx}
\usepackage{subcaption}
\usepackage{booktabs}

\usepackage{tikz}
\usepackage{lipsum}

\begin{document}

\title{A Syndrome-Extraction Framework for Distributed Lattice Surgery on Arbitrary Rotated Surface-Code Layouts}

\author{Pritesh Thakur}
\affiliation{Brandeis University, Waltham, MA, 02453, USA}
\affiliation{Mathematics and Computer Science Division, Argonne National Laboratory, Lemont, IL 60439, USA}
\email{pritesht@brandeis.edu}
\orcid{0009-0008-2391-1911}

\author{Daniel Dilley}
\affiliation{Mathematics and Computer Science Division, Argonne National Laboratory, Lemont, IL 60439, USA}
\email{ddilley@anl.gov}
\orcid{0000-0002-8821-4059}

\author{Zain Saleem}
\affiliation{Mathematics and Computer Science Division, Argonne National Laboratory, Lemont, IL 60439, USA}
\email{zsaleem@anl.gov}

\maketitle

\begin{abstract}
Modular quantum computing is a leading candidate for scalable fault tolerant quantum computation. Performing lattice surgery using surface code patches in different modules requires the use of inter-module gates that are noisier than the local gates. Furthermore, the merged layouts during lattice surgery expose multiple directions in which hook errors can propagate to reduce the fault distance. Existing hook-avoiding schedules either increase the syndrome extraction circuit depth, or let link faults propagate into the bulk. We introduce a seam construction that allows splitting arbitrary merged rotated surface code layout into smaller patches with uniform boundaries that could keep their own hook-avoiding N/Z shaped schedule. At the seam we split weight-4 stabilizers into pairs of weight-2 stabilizers that are mediated through a Bell pair and measured within a CNOT depth of 4. We also employ the use of three-qubit $\texttt{CXX}$ gates for parity measurements at the seam to maintain the minimum CNOT depth. Through this seam construction, we confine the elevated link error at the seam, protecting the bulk from higher error rate. We also give a custom detector annotator for such seams that can find the detector space, annotate detectors in the Stim circuits, and ensure that they are suitable to use by matching decoders. We implement our construction for the logical Pauli-$XX$ measurement lattice surgery and a cross-shaped spatial junction and sweep the link error rate from $10$ times to $100$ times a fixed bulk error rate, and show that our construction gives the lowest and most stable logical error rate than the other frameworks compared here.
\end{abstract}

\section{Introduction}

Quantum computing promises to solve problems that are highly expensive or infeasible for classical computers \cite{feynman1982,lloyd1996universal,arute2019quantum}. As the quantum devices scale beyond the Noisy Intermediate-Scale Quantum (NISQ) era, quantum error correction (QEC) becomes very essential \cite{Preskill_2018, shor1995scheme}. Realizing QEC with tolerable overhead is a central challenge to attain to fault tolerance \cite{Fowler_2012,Campbell_2017}. For this, several quantum error correcting codes have been proposed and constantly modified for improved performance \cite{Kitaev_2003,Bombin_2006,google2024}. 

Surface codes are the current gold standard as they offer higher threshold with only nearest-neighbor interaction and are compatible with planar hardware architecture \cite{Dennis_2002,Fowler_2012,google2024}. Yet, a major issue with surface codes is the qubit overhead. A $d$ distance rotated surface code uses $2d^2 - 1$ physical qubits to encode a single logical qubit \cite{Bombin_2006, Tomita_2014, orourke2024comparepairrotatedvs}. The code distance must be increased to suppress the logical error rate to a level required for useful quantum computation~\cite{google2023suppressing}. However, hardware restrictions limit the number of physical qubits a single quantum processing unit (QPU) can host. For example, fan-out wiring in silicon spin qubits \cite{vandersypen2017interfacing}, spectral crowding of motional modes in trapped ion qubits \cite{cetina2020quantumgatesindividuallyaddressedatomic}, cryostat size and chip fabrication in superconducting qubits \cite{ang2022architecturesmultinodesuperconductingquantum}, photon loss and probabilistic entangling gates in photonic qubits \cite{knill2001scheme, rudolph2017optimistic}, finite laser power and microscope field of view in Rydberg arrays \cite{Saffman_2016, Saffman_2019}, and so on, limit the number of physical qubits a single QPU can contain. 

These hardware constraints imply that a single monolithic quantum processor cannot host an arbitrarily large surface code. Thus, modular quantum computing architecture is considered the leading candidate for scalable fault tolerant quantum computing \cite{monroe2014, ramette2023, aghaee2025scaling}. Similar modular and networked architectures are also essential for the quantum internet and long-distance entanglement distribution \cite{kimble2008}, secure blind and delegated quantum computation \cite{Broadbent_2009, Barz_2012}, distributed quantum sensing and metrology such as networks of atomic clocks \cite{K_m_r_2014, Guo_2019}, and distributed quantum algorithms that pool the resources of separate processors \cite{Yimsiriwattana_2004, xiao2022}.

Distributing a single rotated surface code across different QPU raises the problem of performing non-local inter-module check operations, i.e., performing a parity check on data qubits that lie on one QPU using an ancilla that lies on another QPU. A widely used solution for this problem is to use joint measurement outcomes from shared entangled pairs \cite{monroe2014, ramette2023, Jacinto_2026}. In existing modular architectures, multiple copies of smaller modules are joined together with a noisy link to form a larger network of quantum processors. Here, we define \textit{bulk noise} to be the noise caused by local single or multi qubit gates within the surface code in the same QPU, and  \textit{link noise} to be the noise caused by non-local operation across different QPUs. It has been established that even with the link noise being an order of magnitude higher than the bulk noise, the quantum computation can still be performed fault tolerantly \cite{ramette2023}.

Lattice surgery is increasingly considered a standard primitive for performing fault tolerant logical operations between surface code patches \cite{Horsman_2012, Litinski_2019}. It has also been demonstrated experimentally \cite{Erhard_2021, wang2026, besedin2026lattice}. It is considered better than braiding defects or traversal gates because it preserves the physical qubit and spatial overhead for logical operations, and it allows us to entangle logical qubits through a simple merge and split procedure, using only nearest neighbor interactions and planar connectivity \cite{Horsman_2012, Litinski_2019}. For example, a joint $\overline{XX}$ or $\overline{ZZ}$ parity measurement is performed by merging two patches: stabilizer checks at the shared boundary are activated for $d$ syndrome extraction rounds to measure the joint logical operator fault tolerantly, after which the patches are split apart again \cite{Litinski_2019}. A logical CNOT can be assembled from these joint measurements after the ancilla patch is measured out and recorded. In a modular setting, the two patches may live on different QPUs, which makes the seam exactly the location where the non-local inter-module checks act. So, lattice surgery techniques that can assist fault tolerant surgery even at elevated inter-module noise are required. Existing methods either increase qubit overhead, increase syndrome extraction circuit depth, or result in distance reducing hook errors.

\begin{figure}[t]
    \centering
    \includegraphics[width=0.95\linewidth]{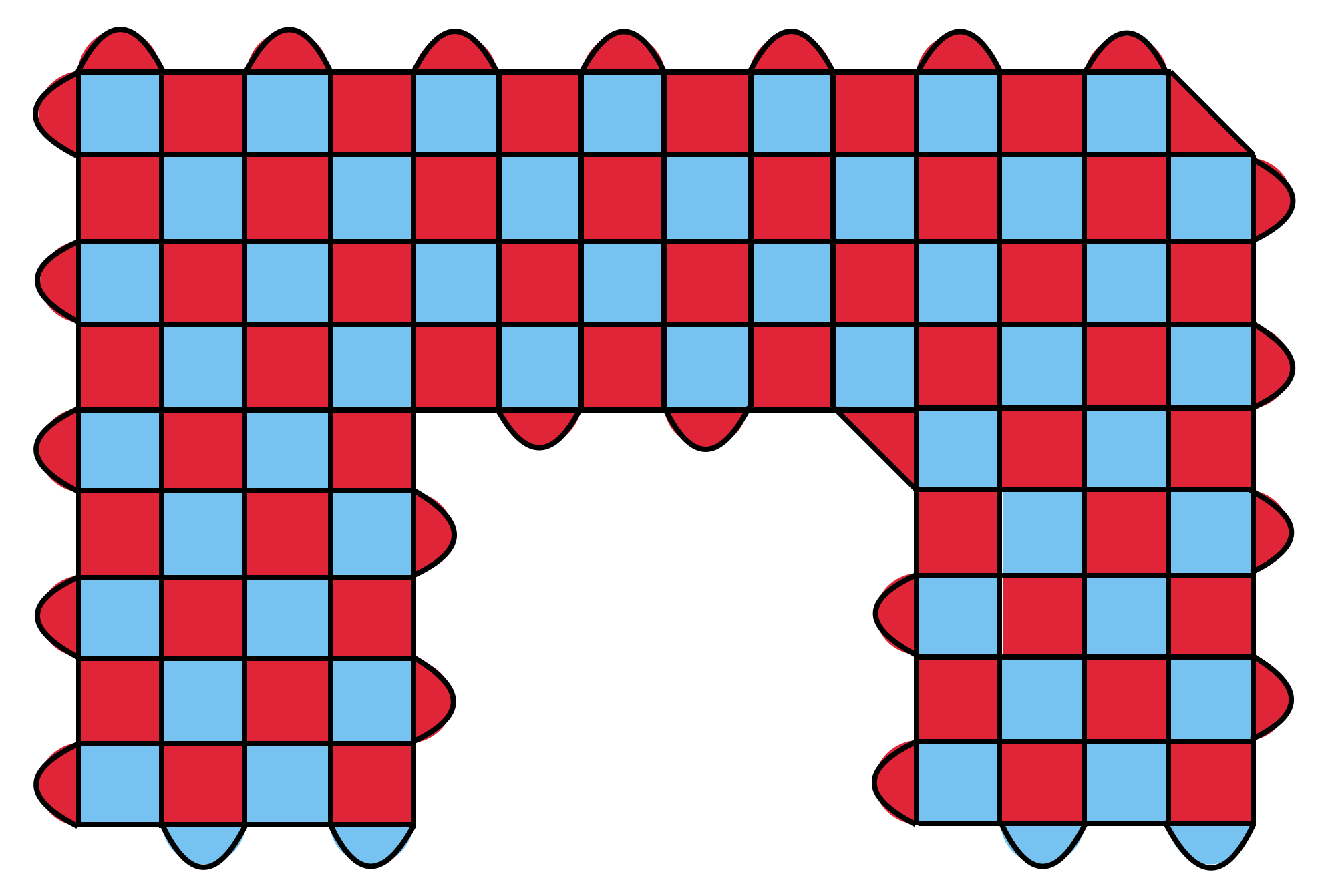}
    \caption{Surface code layout for using lattice surgery to perform logical Pauli-XX measurement between two distance surface code patches. Red (blue) plaquettes denote X (Z) type stabilizers.}
    \label{fig:pauliXX-merge}
\end{figure}

\begin{figure}[t]
    \centering
    \includegraphics[width=0.95\linewidth]{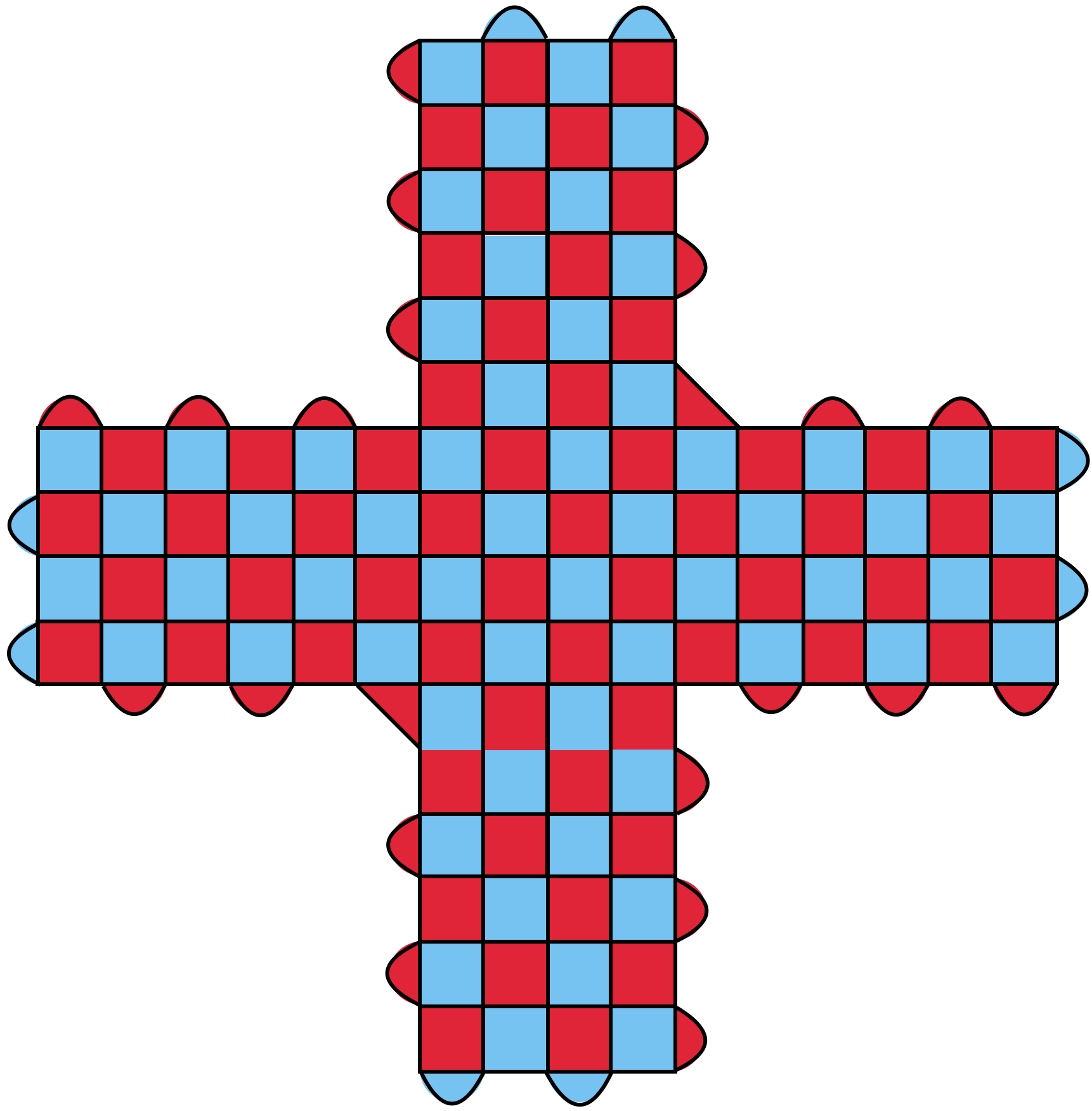}
    \caption{Surface code layout for Cross-shaped spatial junction merge. Red (blue) plaquettes denote X (Z) type stabilizers.}
    \label{fig:x-merge}
    
\end{figure}

In this work, we introduce a seam construction for modular lattice surgery that lets each merged patch retain its own hook-error-avoiding syndrome extraction schedule without increasing the syndrome extraction circuit depth and substantial physical qubit overhead. We demonstrate our construction by performing lattice surgery for logical Pauli-$XX$ measurement between two distant surface code patches (Figure~\ref{fig:pauliXX-merge}) and a Cross-shaped spatial junction merge (Figure~\ref{fig:x-merge}). We show that our construction gives lower logical error rates compared to the benchmark techniques discussed in section~\ref{sec:prior}.

The remainder of the paper is organized as follows. Section~\ref{sec:problem} introduces and formalizes the scheduling problem that hook errors cause at modular seams and also in the bulk of surface code with unconventional boundaries. Section~\ref{sec:prior} reviews existing constructions for avoiding hook errors and inter-module lattice surgery frameworks. Section~\ref{sec:construction} explains the circuit construction and detector annotation for our framework. Section~\ref{sec:simulation} presents our simulation methodology, and Section~\ref{sec:results} presents our results and compares the performance of the different frameworks. We conclude and explain some future directions in Section~\ref{sec:outlook}.

We assume that the reader has a basic understanding of quantum error correction, surface codes, detectors, logical operations, and lattice surgery. Although in syndrome extraction circuits, Controlled-$X$ (CNOT or also referred as \texttt{CX}) or Controlled-$Z$ (\texttt{CZ}) gates could be used to extract parity information, in this paper we use the term ``CNOT depth'' to mean the number of steps of these two-qubit gates used during a single QEC syndrome extraction cycle. The minimum CNOT depth for a standard N/Z shaped syndrome extraction schedule is 4 during merge operations that do not cause hook errors.

\begin{figure}[t]
    \centering
    \includegraphics[width=0.95\linewidth]{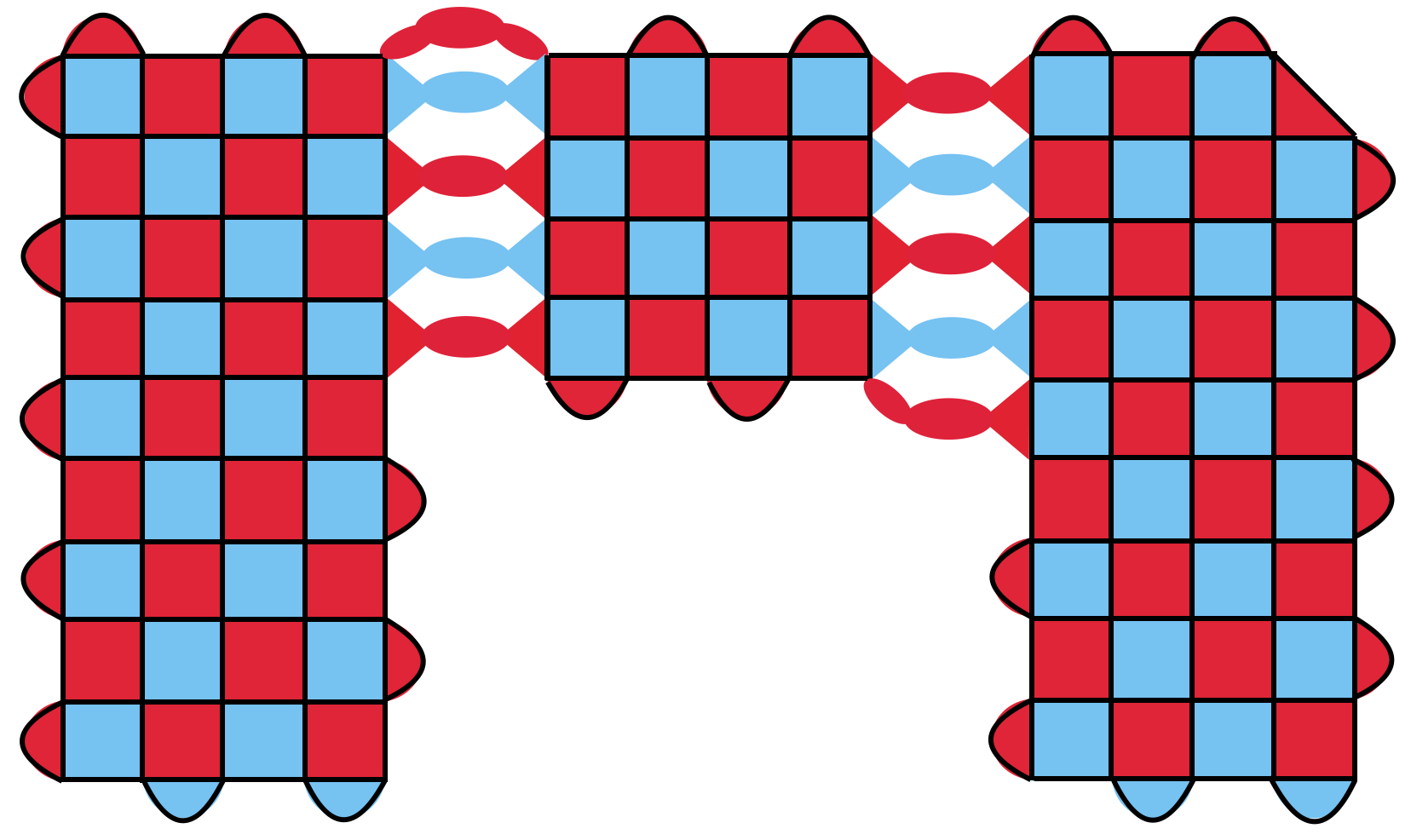}
    \caption{Merged surface-code layout for a logical Pauli-$XX$ measurement. Red (blue) plaquettes denote X (Z) type stabilizers. Each ancilla at the seam is split into two two-body checks. This allows us to join patches of differing boundary structure across a module boundary.}
    \label{fig:pauliXX_bellsplit}
\end{figure}

\section{Problem Setting}
\label{sec:problem}

Lattice surgery is essential in surface code-based fault-tolerant logical quantum computation. It involves merging and splitting patches with different boundary structures from different directions. Such a merge changes the boundary structure of the participating patches and, with it, the constraints that a fault-tolerant syndrome extraction schedule must satisfy. Thus, creating a syndrome extraction schedule for lattice surgery on arbitrary rotated surface code layouts while ensuring protection from distance-reducing hook errors is an important and non-trivial task. In a modular setting, the schedule must additionally minimize circuit depth to reduce the accumulation of idling errors while preventing link noise from significantly degrading the overall logical error rate.

In a standard rotated surface code, each interior syndrome qubit measures the parity of its four adjacent data qubits, whereas each edge syndrome qubit measures the parity of its two adjacent data qubits. These parity check operations must be performed in a proper order so as to satisfy the following criteria \cite{orourke2024comparepairrotatedvs}:

\begin{itemize}
    \item \textbf{Determinism and Commutativity:} The order of the two-qubit gates for syndrome extraction must be set to ensure that the parity information obtained is deterministic. For example, let stabilizer A and stabilizer B share two data qubits $\ket{q_1}$ and $\ket{q_2}$. If stabilizer A applies its two-qubit gate to $\ket{q_1}$ before stabilizer B applies a two-qubit gate to either shared qubit, then stabilizer A must also apply its two-qubit gate to $\ket{q_2}$ before stabilizer B does so.

    \item \textbf{Parallelism:} The syndrome extraction schedule for the two types of checks must be on a parallel axis. This prevents collisions between two-qubit gates. A single data qubit can participate in only one two-qubit gate with a check qubit at a particular time step.

    \item \textbf{Avoid hook error:} For rotated surface codes, the syndrome extraction schedule for a stabilizer check must be performed such that the last two data qubits that the two-qubit gates operate on do not align with the direction of the logical operator corresponding to the stabilizer whose check is performed. If not, a single physical error gets copied to two data qubits, reducing the effective distance of the rotated surface code \cite{Dennis_2002}.
\end{itemize}

The merging of rotated surface-code patches in an arbitrary layout can expose several directions along which distance-reducing hook errors propagate. For example, in the merged surface code layout of the logical Pauli-$XX$ measurement shown in Figure~\ref{fig:pauliXX-merge}, a hook running either top-to-bottom or right-to-left shortens the $X$-fault distance. In the cross-shaped spatial junction of Figure~\ref{fig:x-merge}, which direction is dangerous depends on the arm: right-to-left hooks reduce the $X$-fault distance on the top and bottom patches, while top-to-bottom hooks reduce it on the left and right patches.

To let each patch keep its own hook-avoiding N/Z shaped syndrome extraction schedule, one might naively split the merged region into separate patches along their differing boundaries. For the layouts shown in Figures~\ref{fig:pauliXX-merge} and~\ref{fig:x-merge}, this leaves three patches to be joined end to end, with a seam bridging each adjacent pair. Usually, a seam check across a module boundary is performed by splitting a single weight-4 stabilizer into two weight-2 stabilizers. This leaves one ancillary qubit on each side of the seam as shown in Figure~\ref{fig:pauliXX_bellsplit}. Doing so introduces two problems.

First, splitting the ancilla reorients the seam checks, opening more pathways for hook errors. To satisfy the determinism and parallelism criteria for the syndrome extraction schedule, we need to allow each half to retain the syndrome extraction schedule of the bulk to which it is attached. Now, the $X$ checks at the seam act vertically along the seam. This causes a single physical qubit error to be copied onto two data qubits aligned with the seam. Successive checks chain these hooks into an error string running the length of the seam. This string lies along the top to bottom direction identified above as reducing the $X$-fault distance. So, the split reintroduces the hooks that the per-patch schedules were meant to solve. 

Second, joining three patches end-to-end requires two such seams. The center patch must merge with two patches on both sides. Satisfying the above listed criteria while still letting each patch keep its own hook-avoiding schedule at minimum CNOT depth puts heavy constraints on the parity checks at the center patch. The two edges associated with the seam impose orderings that cannot be satisfied within the minimum CNOT depth. We resolve both obstacles in Section~\ref{sec:construction}, where each patch retains its own hook-avoiding schedule across both seams at CNOT depth of 4.

\section{Existing Work}
\label{sec:prior}

In this section, we review some existing constructions for avoiding hook errors and merging patches across modules, and build the setup to explain our construction.

A global temporally-alternating syndrome extraction schedule was introduced in Ref.~\cite{gidney2025alternating,Bluvstein_2025}. The idea is to alternate the ancilla's CNOT ordering between rounds of syndrome extraction, for example, alternating between ``N'' and ``Z'' shaped CNOT orderings. This sends hook errors in perpendicular directions so that the hook-prone order occurs in every other round. It gives the fault distance of $d-1$. However, this schedule is imposed uniformly across the entire patch, and a single global ordering might not remain hook-avoiding for the merged patch. For example, both the N- and Z-shaped schedules for $X$-checks remain hook-prone in the top-right and top-left portions of the surface code layout shown in Figure~\ref{fig:pauliXX-merge}. Additionally, when applied in a modular setting, it performs the inter-module coupling in the interior of the extraction cycle. Therefore, link errors propagate into the bulk, and the logical error rate degrades sharply with elevated inter-module noise. 

Ref.~\cite{Litinski_2018} spatially varied the N/Z shaped schedule through the code. Avoiding gate collisions increases the CNOT depth for this method, and it will also fall into the same problem of logical error rates sharply increasing in elevated link noise setting that Ref.~\cite{Bluvstein_2025} falls into.

Ref.~\cite{kishony2026} introduced the ``diagonal schedule'' for syndrome extraction. Here, the parity checking two-qubit gates for each plaquette are ordered such that the last two data qubits being checked lie on a diagonal of the plaquette. This makes hook errors propagate along the plaquette diagonal, which never aligns with a horizontal or vertical logical operator. So, the full code distance $d$ is preserved irrespective of the boundary geometry. Using this method, a CNOT depth of $4$ is attained, but only on hardware that can execute qubit reset and measurement at the same time step as the two-qubit gates on other qubits. On most platforms, resets and measurements are comparable in time duration to that of entangling gates \cite{google2023suppressing} and cannot overlap with them. Where overlap is not allowed, the period grows to eight or nine steps~\cite{kishony2026}, increasing the effective physical error rate per cycle.

The idea of dynamically moving stabilizer tiles, using degree-3 connectivity, was introduced by Ref.~\cite{McEwen_2023}, where the $Z$ tiles drift toward the $Z$ boundary, and the $X$ tiles drift toward the $X$ boundary each round. Ref.~\cite{hirai2026} observed that this tile motion shortens the hook-error edges in the decoding graph and repurposed it as a hook-avoiding syndrome-extraction scheme. Here, each measurement qubit couples to only three neighboring data qubits, the cycle retains the minimum CNOT-depth of $4$, and retains the full fault distance $d$. However, depending on the layout of the lattice surgery merge, the drifting tiles should be forced towards boundaries of the opposite Pauli type. This increases the physical qubit overhead at the boundary. Moreover, though it eliminates the problem of hook errors, this framework is targeted at monolithic quantum computation. When it is applied in a modular setting, the traveling tiles carry the link fault to the bulk as a propagating data error rather than confining it to a single measurement outcome. So, its logical error rates rise sharply as the inter-module noise increases (Section~\ref{sec:results}).

Ref.~\cite{Shalby_2025} compares three strategies for connecting two surface-code patches across a noisy inter-module link: a bare direct link (DL), a cat-state (CAT) gadget~\cite{Huang_2021}, and a gate-teleportation (GT) gadget~\cite{Gottesman_1999}. They compare these three frameworks in both rotated and unrotated surface codes, with the boundary-crossing gates up to ten times noisier than the bulk. After careful optimization, all three frameworks give the full fault distance $d$. But for CAT and GT gadgets, this is achieved at the cost of increasing the CNOT depth to $5$ and using additional physical qubits. They claim DL to be the best framework as it operates with a minimum CNOT depth of $4$ without additional physical qubits, and it gives the lowest logical error rates compared to the others. However, these interfaces are tested only in quantum-memory experiments between two patches, not in lattice surgery with multiple patches joining end to end. Therefore, we take DL as our baseline. DL keeps the standard syndrome-extraction circuit and treats the boundary-crossing CNOTs as inter-module gates with an elevated link error rate. Since DL leaves the CNOT schedule unspecified, and lattice surgery additionally requires the schedule to avoid hook errors, we instantiate DL with two hook-avoiding schedules: the interleaving schedule of Ref.~\cite{hirai2026} and the mirror-alternating schedule of Ref.~\cite{haug2025latticesurgerybellmeasurements}, described below.

Ref.~\cite{Jacinto_2026} introduced measurement teleportation. They split a single 4-body stabilizer into two parts: 3-body and 1-body stabilizers. Their work flow is as follows: prepare the split ancilla in a bell state, perform parity check operations, measure both in the computational basis, and finally combine the results by XORing the individual outcomes, and use the combined outcome for error correction. This work is very close to our setting and solves the same problem that we solve. It achieves full fault distance $d$ but at the cost of using $2d$ bell pairs per inter-module seam. We achieve a fault distance of $d-1$ using only $d$ bell pairs per seam. Also, this work treats a Bell state as an external resource and does not show the bell state generation step, which could further introduce data qubit idling. We choose not to compare the performance of our framework against this work because this poses different hardware architecture requirements than us, as shown in Figures 2, 4 of Ref.~\cite{Jacinto_2026}.

One of our two seams uses the Bell-measurement idea proposed in Ref.~\cite{haug2025latticesurgerybellmeasurements}. A single four-body stabilizer check is split into two two-body stabilizers, each split ancilla keeps the same schedule as the bulk, and the Bell-creating inter-module CNOT is performed as the last two-qubit gate of the syndrome extraction cycle. After this CNOT gate, at the ancilla measurement step, the control ancilla is measured in the $X$-basis and the target in the $Z$-basis, and the XOR of the two outcomes is used for error correction. To avoid hook errors at the seam, the bulk CNOT ordering alternates between rounds, not between N- and Z-shaped orderings as in Refs.~\cite{gidney2025alternating,Bluvstein_2025}, but between an ordering and its mirror image, so that the hook-prone pair of data qubits moves between opposite edges of the plaquette from round to round. We refer to this as the \emph{mirror-alternating schedule}. This construction attains a CNOT depth of $4$ only in the situation when two patches are merged together. Merging three patches end to end, as required here, increases the CNOT depth. We resolve this in Sec.~\ref{sec:construction}. Ref.~\cite{hirai2026} uses the mirror-alternating schedule without the Bell-measurement as a baseline for their interleaving schedule, and in our section~\ref{sec:results}, we present a comparison of our framework against this mirror-alternating schedule without the bell seam.

\section{Our Framework}
\label{sec:construction}

We propose a syndrome extraction framework that simultaneously (i) avoids the pathological distance-halving hook error chains, (ii) gives stably low logical error rates even when the link noise is elevated by an order of magnitude, (iii) allows us to split the surface code layout into smaller patches that can keep their own N/Z shaped schedule at minimum circuit depth, and (iv) enables us to join the smaller patches together such that the logical structure of the merged patch is preserved. We designed the circuits and surface code layouts in Crumble and tested them in Stim \cite{crumble, gidney2021stim}.  

Before going through the framework, we introduce some terms used here. We partition a large patch of surface code into smaller regions with uniform boundaries that can be distributed across different QPUs. To join these regions, we take a column of ancilla and split each of them into halves, changing a single four-body stabilizer into pairs of two-body stabilizers. We often refer to them as \textit{split-ancilla}, and we refer to the column where the split occurs as the \textit{seam}. Throughout, we refer to the two-qubit gates that act between two qubits within the same module as \textit{bulk gates} that are subject to \textit{bulk noise}, and we refer the two-qubit gates that cross the module to act on qubits that lie on different modules as the \textit{inter-module gates} that are subject to \textit{link noise}.

\begin{figure}
    \centering
    \includegraphics[width=1\linewidth]{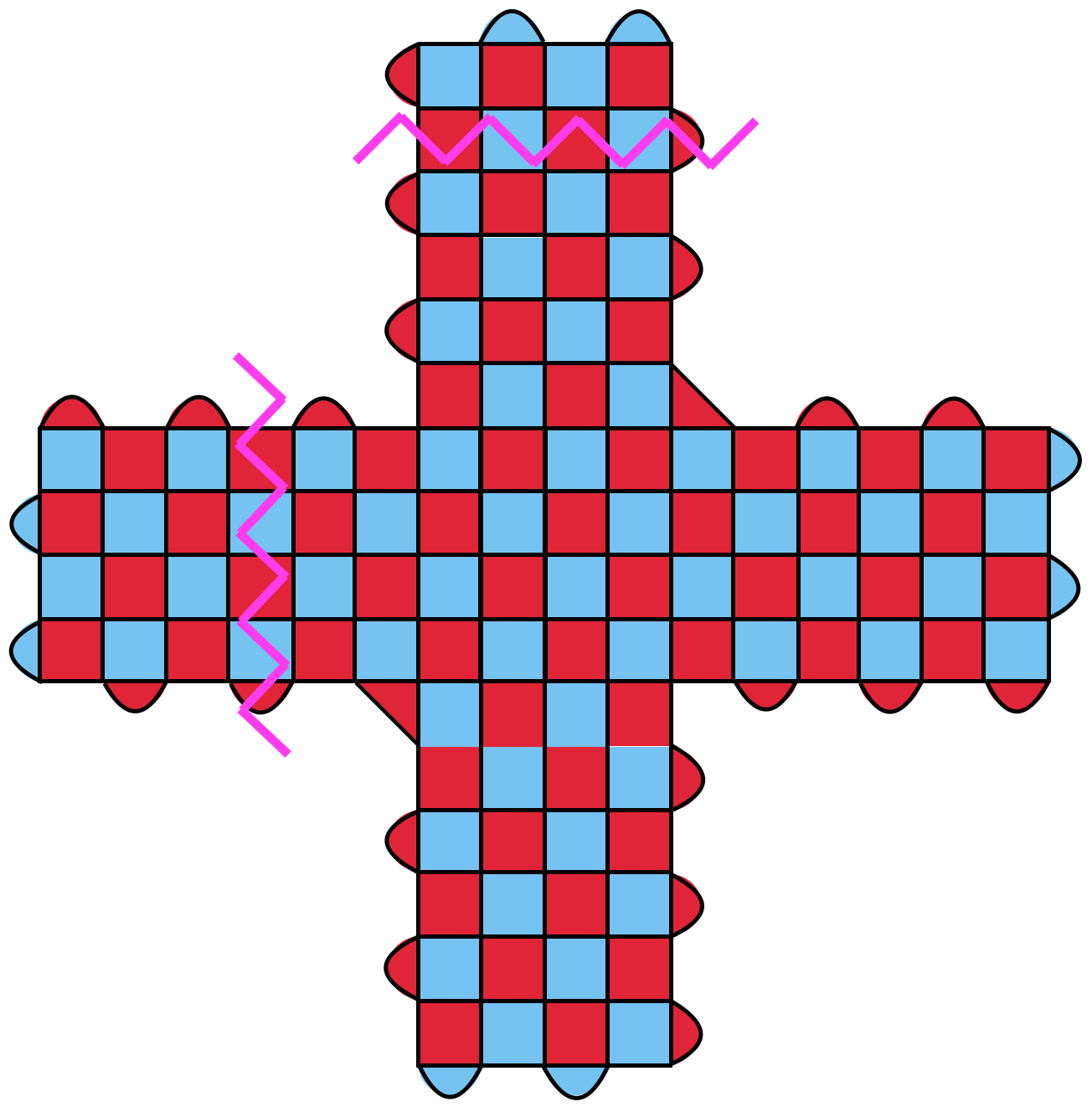}
    \caption{In the most pessimistic assumption, the two magenta chains in this picture reflect the two possible directions from which the hook-induced logical-X error chains could occur during the crossing-shaped junction merge. Such X-type hook error chains halve the distance of the Z-logical observable that is being observed during this lattice surgery merge.}
    \label{fig:x-merge-hooks}
\end{figure}

\begin{figure}[t]
    \centering
    \includegraphics[width=1\linewidth]{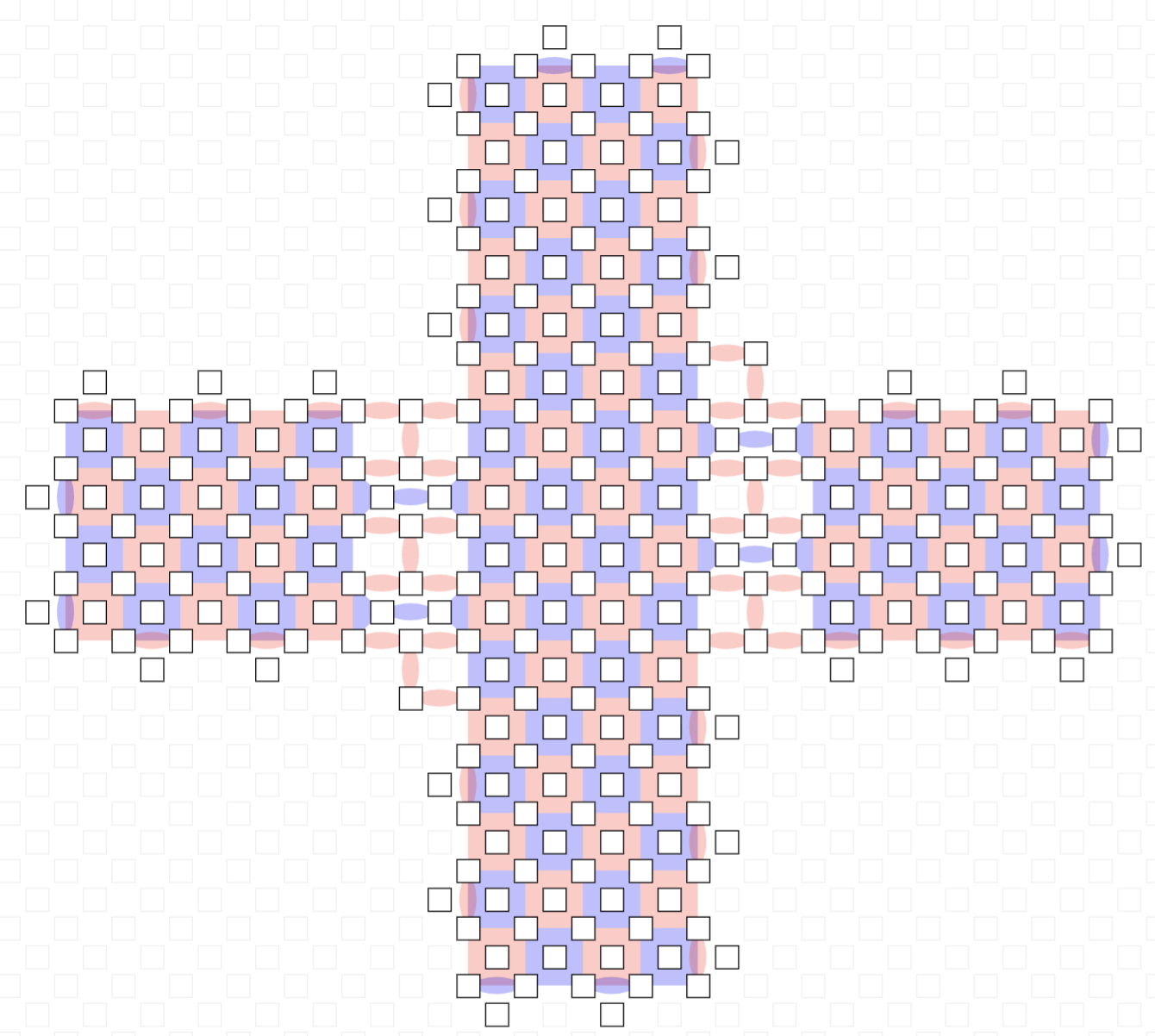}
    \caption{CRUMBLE circuit showing the cross-shaped spatial junction merge layout being split into three patches that can keep their own hook-avoiding N/Z shaped schedule at minimum circuit depth. Red polygons denote X-type stabilizer checks and blue polygons denote Z-type stabilizer checks. The Z-type and X-type stabilizers are split differently at the seam, as reflected by the blue and red polygons at the seam.}
    \label{fig:x-merge-fix}
\end{figure}

\begin{figure*}
    \centering
    \includegraphics[width=\textwidth]{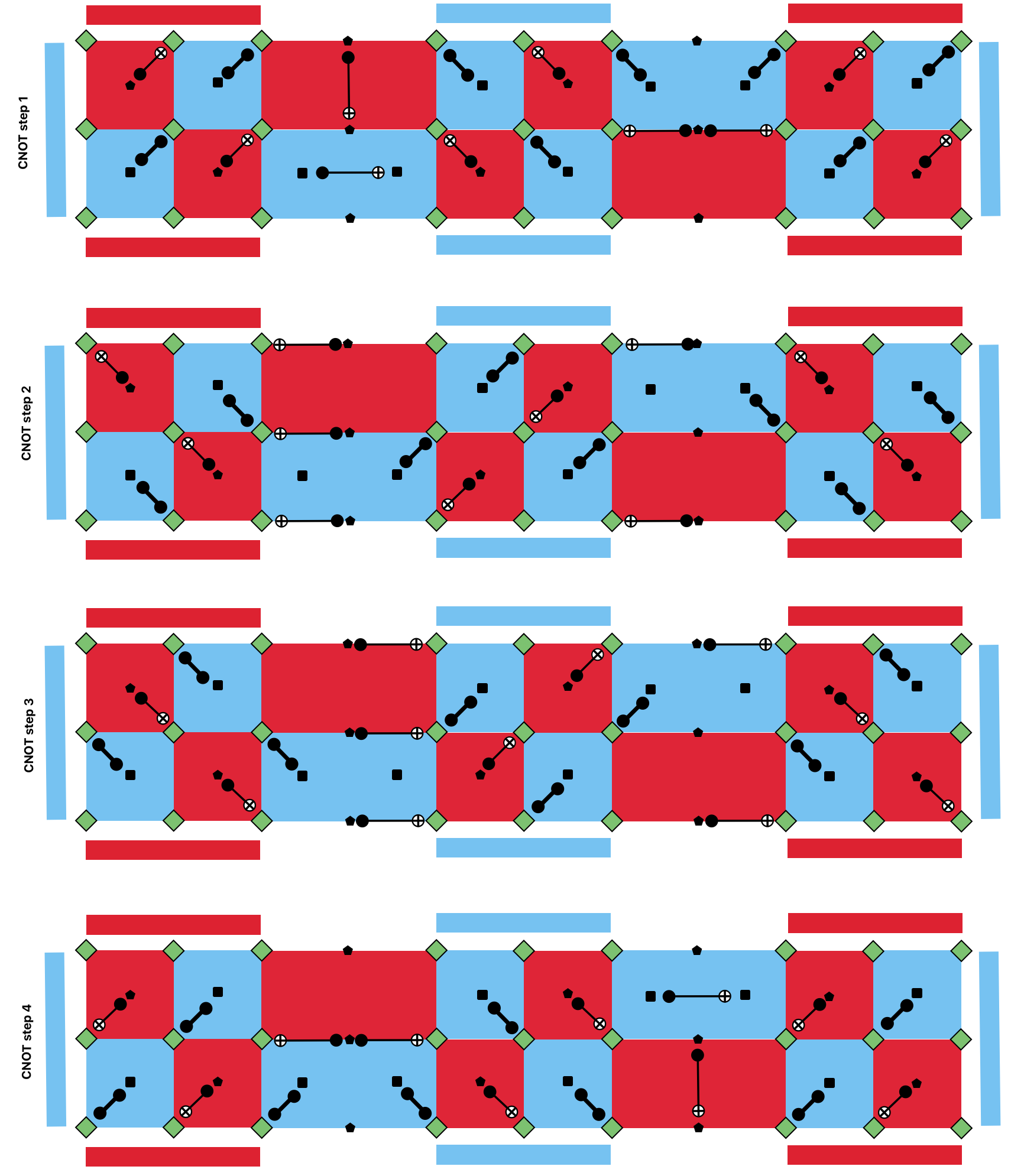}
    \caption{Syndrome extraction framework joining three surface code patches with different boundaries, connected end to end. Red polygons denote X-type stabilizers and blue polygons denote Z-type stabilizers. Color-filled rectangles are placed out of the main surface code patch to reflect the boundary type of each patch: red-filled rectangles denote to X-type boundary and blue-filled rectangles denote to Z-type boundary. Green-filled diamond shapes denote the data qubits. Black-filled squares at the center of blue tiles denote Z-measurement qubit, and black-filled pentagons at the center of the red tiles denote X-measurement qubit. The standard notations for CNOT and CZ gates are used to denote the respective parity checking two-qubit gates for X-type stabilizers and Z-type stabilizers.}
\label{fig:cx-steps}
\end{figure*}

Our framework was developed in an effort to merge three surface code patches with different hook directions, end-to-end, while still preserving the logical structure of the full code. First, we explain our work by taking the cross-shaped spatial junction merge layout (Figure~\ref{fig:x-merge}) as an example. Figure~\ref{fig:x-merge-hooks} shows the directions that need protection from X-type hook errors. Note that when we look for the length of shortest graph-like error for the logical $Z$-observable, we are looking for the shortest error chain of logical-$X$ errors that anti-commute with that $Z$-observable. In Figure~\ref{fig:x-merge-hooks}, the central region needs protection from $X$-type hooks in the horizontal direction, and the right and left arms need protection from the $X$-type hooks in the vertical direction.

A naive solution to avoid these $X$-type hook errors would be to use different schedules for the $X$-type stabilizers: use an N-shaped schedule at the central region and a Z-shaped schedule on the right and left arms of the layout. At the same time, the schedule for $Z$-type stabilizers must also be hook avoiding. So, for the $Z$-type stabilizers, we need a $Z$-shaped schedule in the central region and an N-shaped schedule in the left and right arms. Although this looks like an easy fix, implementing these checks while still satisfying the basic criteria for the syndrome extraction schedule listed in section~\ref{sec:problem} forces the CNOT depth to increase to 7 from the minimum CNOT depth of 4, inviting a lot more physical qubit idling.

In our framework, we divide the layout into three patches: left, center, and right. Figure~\ref{fig:x-merge-fix} shows the layout of our construction. The blue polygons denote $Z$-type stabilizers and the red polygons denote $X$-type stabilizers. We split the $X$-type stabilizers and the $Z$-type stabilizers differently at the seams. As shown in figure~\ref{fig:x-merge-fix}, we split the $Z$-type stabilizers vertically into two halves and the $X$-type stabilizers horizontally into two halves.

\subsection{Gate Schedule}

It is not possible to explain our CNOT scheduling method fully in a single time frame. Instead, we give a step-by-step explanation of how it works. The four CNOT steps of our syndrome extraction framework are clearly presented in the four time frames of Figure~\ref{fig:cx-steps}, where the red and blue polygons still retain their same meaning. 
The image shows three $3\times3$ surface code patches that are joined end-to-end. The blue and red rectangles on the right and left of the central logical patch represent the seam that we use to join the three patches together, preserving the logical structure of the full merged code. For the sake of clarity, we use the standard symbols for CNOT and CZ gates that are used to check the parity of the data qubits: CNOT gates are used to perform stabilizer measurements for $X$-type stabilizers and CZ gates are used to perform stabilizer measurements for $Z$-type stabilizers.

Figure~\ref{fig:cx-steps} can be treated as a zoomed in version of the cross-shaped spatial junction merge shown in Figure~\ref{fig:x-merge-fix}. In Fig.~\ref{fig:x-merge-fix}, the left and right arms have X-type boundaries on the top and bottom, Z-type boundaries on the sides; the central patch has Z-type boundaries on the top and bottom and X-type boundaries on the sides. In Fig.~\ref{fig:cx-steps}, we use color-filled rectangular shapes that do not lie on the main patch to denote the boundary types: red-filled rectangles denote the X-type boundaries and blue-filled rectangles denote the Z-type boundaries. 

In the \emph{bulk} of the right and left patches of Figure~\ref{fig:cx-steps}, we use the Z-shaped schedule for the $X$-type stabilizers and N-shaped schedule for the $Z$-type stabilizers. In the \emph{bulk} of the central patch, we use N-shaped schedule for the $X$-type stabilizers and Z-shaped schedule for the $Z$-type stabilizers. These shapes for the schedules can also be clearly seen by following the four time frames of Figure~\ref{fig:cx-steps}. For an additional layer of protection against hooks, we alternate between these schedules and their mirror opposite schedules~\cite{haug2025latticesurgerybellmeasurements}. The alternating schedule provides protection against hook errors when the two possible hook directions lie on the same surface code patch. As an example, consider Figure~\ref{fig:pauliXX-merge}, where the top-left and top-right regions can have hook error chains running from top to bottom and right to left. 

At the seam, the $Z$-type stabilizers are split into pairs of two-body stabilizers vertically. So, each patch has its own two-body stabilizer ancilla. Each ancilla is assigned the same check schedule as the $Z$-type stabilizer schedule in the bulk of the surface code patch to which they are attached. This prevents gate collisions and helps us maintain a CNOT depth of $4$. 

To further eliminate the problem of the distance-halving hooks at the seam, we split the $X$-type stabilizers horizontally. Each of the two-body stabilizers thus check two data qubits, one from each of the surface code patches neighboring the ancilla. Unlike the $Z$-type stabilizers at the seam that only check the data qubit of the patch to which they are attached, the $X$-type stabilizers at the seam perform parity measurements on the data qubits located in both patches.

While implementing this schedule, we run into a situation where a single ancilla needs to perform a parity check operation on two neighboring data qubits at the same time step to maintain the minimum CNOT depth to $4$. For example, consider the CNOT step 1 for the left seam and CNOT step 4 for the right seam of figure~\ref{fig:cx-steps}. In this scenario, we use a three-qubit $\texttt{CXX}$ gate to perform the joint parity measurement of the two data qubits using the single syndrome qubit as the control. In the Figure~\ref{fig:cx-steps}, this three qubit $\texttt{CXX}$ gate has been shown as two separate CNOT gates for clarity. The three-qubit $\texttt{CXX}$ gate required at this step is not a demanding hardware primitive. For superconducting qubits, Ref.~\cite{tasler2025optimizingsuperconductingthreequbitgates} designed and numerically optimized a $\texttt{CZZ}$ parity checking gate, which is equivalent to our $\mathrm{CXX}$ up to single-qubit basis changes on the data qubits. It is performed as two $ZZ$ interactions applied in parallel through tunable couplers. Thus, it can be applied in a single time step and does not impose a qualitatively new requirement on the hardware. Ref.~\cite{tasler2025optimizingsuperconductingthreequbitgates} reported a simulated gate duration of $35\,\mathrm{ns}$ and fidelity of $99.96\%$, and Ref.~\cite{Old_2025} showed that stabilizer-readout circuits based on such three-qubit gates can be made strictly fault-tolerant. Additionally, using multi-qubit entangling gates for quantum error correction is a growing research direction as shown in neutral-atom arrays~\cite{Evered_2023}, trapped ions~\cite{Lu_2019}, and superconducting qubits~\cite{Xu_2026}.

We implement distinct protocols at our right and left seams. Our effort in this framework is to keep all the two-qubit operations within the four CNOT steps of syndrome extraction, and that includes the inter-module CNOT gates performed to prepare the split ancilla into bell states or to perform bell measurements. Following the above syndrome extraction schedule for the bulk and split ancilla, as well as needing to avoid gate collisions, we are forced to perform inter-module CNOT gates on the split ancilla at different time steps for the right and the left seams.

As shown in Figure~\ref{fig:cx-steps}, the inter-module CNOT operation is performed at step 1 for the left seam, but at step 4 for the right seam. The split ancilla at the left seam are thus prepared in a Bell state prior to the syndrome extraction steps. Each of them are then used to perform their local parity checks and finally both are measured in the $X$ basis. Their joint parity information is used for decoding. At the right seam, the inter-module CNOT gate is performed at the last CNOT step after the split ancilla have finished doing their local parity check operations. Then, the control ancilla of the inter-module CNOT gate is measured in $X$ basis and the target is measured in $Z$ basis, and their joint parity information is used for decoding. Thus, the left seam can be treated as following the ``measurement teleportation'' protocol introduced in Ref.~\cite{Jacinto_2026}.

Instead of taking the Bell state from some external source, which could result in idling of the data qubits, we prepare the Bell state within the syndrome extraction steps, and we use only $d$ bell pairs to carry out the merge operation compared to the $2d$ bell pairs used in Ref.~\cite{Jacinto_2026}. The right seam can be treated as following the bell-measurement protocol introduced in Ref.~\cite{haug2025latticesurgerybellmeasurements}. But note that it is not possible to employ this protocol on the left seam without increasing the CNOT depth. This is our full framework for syndrome extraction in inter-module lattice surgery at minimum CNOT depth, maximum fault distance, and lowest obtained logical error rates while using an elevated link error rate.



\subsection{Detector Annotation}
\label{sec:annotation}

A detector is a set of measurement records whose parity is deterministic in the absence of noise~\cite{gidney2021stim}. Annotating detectors on our seam with bell pairs is a non-trivial task. The flow-based annotator \texttt{tqecd}~\cite{suau2026tqec} was not able to completely annotate our code. It leaves some seam records in no detector at all, and it does not recover all the independent detectors that the seam supports, so the resulting distance falls below the code distance. So, we directly annotate our code from the circuit's measurement statistics. The underlying linear algebra is the same that is used by the flow-based approach, but it acts on a different space. We find the detectors from the space of measurement records rather than the space of Pauli operators. On our merge circuit without Bell-pair seams, our annotator and \texttt{tqecd} produce the same detectors.

\paragraph{Deterministic-parity space.}
Let the circuit contain $N$ measurements. A single noiseless run produces an outcome vector $\mathbf m\in\mathbb F_2^{N}$, whose $i$-th entry $m_i$ is the outcome of the $i$-th measurement in execution order. We refer to this measurement as the \emph{record} $i$.\footnote{These are the measurement records seen in Stim circuit files as \texttt{rec[$\cdot$]} from which detectors are built, e.g.\ \texttt{DETECTOR rec[-1] rec[-62]}. Note that Stim addresses measurements relatively. For example, at a point in the circuit where $M$ measurements have already been performed, \texttt{rec[-k]} denotes the record $i = M-k$ (where $i$ is the i-th index starting from beginning). Therefore, the same offset value refers to different records depending on where the instruction appears. Instead, we use absolute indices in the input and work with them throughout, since the following linear algebra requires a fixed labeling of the coordinates of $\mathbb F_2^{N}$. Remember that this labeling is an arbitrary choice and does not affect the detectors themselves.} 
Sampling the noiseless circuit $T$ times and stacking the outcome vectors as rows gives a matrix $S\in\mathbb F_2^{T\times N}$, whose entry $S_{t,i}$ is the outcome of record $i$ in run $t$.

A detector $D$ is a set of records. It can also be written as an indicator vector
$\mathbf d\in\mathbb F_2^{N}$ with $d_i=1$ for $i\in D$. By definition, the joint parity of these records' outcomes, $\bigoplus_{i\in D}m_i$, stays constant in every noiseless run. The standard reset operations prepare each qubit in a $+1$ eigenstate of a Pauli operator, so this constant parity is $0$ (a $-1$ eigenstate would instead give $1$). Thus, every detector satisfies $S\mathbf d=\mathbf 0$, and the deterministic parities form the null space
\begin{equation}
  \mathcal P \;=\; \ker S \;=\;
  \bigl\{\mathbf d\in\mathbb F_2^{N} : S\mathbf d=\mathbf 0\bigr\},
  \label{eq:parity-space}
\end{equation}
of dimension $\dim\mathcal P = N-\operatorname{rank}S$.

A fault flips a subset of records, breaks the expected parity, and is caught by the detector. Since Eq. \eqref{eq:parity-space} uses only measurement statistics and no knowledge of lattice geometry, it treats bulk, boundary, and seam records alike, so $\mathcal P$ is the complete inventory of detectors available in the circuit. This inventory is an exact property of the circuit, which a flow method computes directly from the Clifford structure of the circuit. The sampling method estimates the same subspace rather than computing it exactly, but the estimate is provably reliable. This is because a genuinely random parity stays constant for $T$ shots with a probability of only $2^{-T}$, and this probability is vanishingly small for $T$ comfortably larger than $N$. Thus, the sampled space coincides with the true space with overwhelming probability.

\paragraph{Discovery.}
We obtain a basis of $\mathcal P$ by performing a Gaussian elimination on the columns of $S$, sweeping the records in measurement order. The column $S_{:,i}$ is the length-$T$ vector of record $i$'s outcomes gathered from the $T$ different shots. As established above, a set of records is a detector precisely when XOR of their columns gives a zero vector. As we sweep through the records, we maintain a set of \emph{live} columns: the records seen so far whose outcome columns are linearly independent, i.e. none of them is the XOR of the others. Each live column is marked with a \emph{pivot}, a shot (a row of $S$) where it reads $1$. When we create a live column, we first cancel it against the existing live columns, so it reads $0$ at each of their pivots and takes a new pivot of its own. The pivots are therefore all distinct, and every live column created later reads $0$ at the earlier pivots. So when we cancel a new column against the live set in order, each pivot bit is cleared once and is never switched back on, which is what makes the cancellation unambiguous.

Records are processed one at a time. For a record $i$, we take its column $S_{:,i}$ and cancel the live columns out from it using their pivots (taking the live columns in order). What remains is the \emph{residue}. If the residue is $\mathbf 0$, record $i$ is fully determined by earlier records. This implies that the records that were XORed together form a deterministic parity, and we emit them as a detector. If the residue is nonzero, it is added to the set of live columns. 
\footnote{A residue that is a constant $1$ rather than $0$ is also deterministic, and it is emitted as a detector as well. This can only happen if the reset is performed in the $-1$ eigenstate. So it does not arise under the standard resets used here.}

Because records are handled in order and canceled only against earlier ones, each detector is supported at or before the record that closes it. This closing record is named the \emph{anchor}. It is the last measurement record that a detector contains. Thus, the parity of the detector is fixed at the moment the anchor is measured. A closing record never re-enters a later combination, so the anchors are distinct and the emitted detectors are independent. There are $N-\operatorname{rank}S=\dim\mathcal P$ of them, and they form a basis of $\mathcal P$. A record that lies in no detector is a \emph{gauge} record. So, a fault that flips a gauge record lights up no detector and passes unnoticed. A vanishing gauge count therefore certifies that every measurement sits in some detector, hence that a fault on any measurement is detectable.

\paragraph{Observable protection.}
A \emph{logical observable} is a set of records whose parity is deterministic and encodes the logical bit. It can also be written as a vector $\mathbf o\in\mathbb F_2^N$. A \emph{logical error} is an error that flips this parity without setting off any detector. Since the logical observable $\mathbf o$ is also deterministic, it also lies in $\mathcal{P}$ alongside the detectors. For an observable $\mathbf{o}$ to be genuine, it should not be reconstructible as an XOR of the detectors. If $\mathbf{o}$ could be constructed as an XOR of the detectors, any error that does not light any detectors could never flip the logical observable. Thus, the observable would not suffer from a logical error and would not have a finite distance. So, $\mathbf o$ must be kept outside the detector span. Therefore, we build the detector span $\mathcal D\subseteq\mathcal P$ greedily. 

We seed $\mathcal{D}$ with the bulk detectors from a separate pattern annotator, and then accept each further discovered parity only if it enlarges the span $\mathcal{D}$ while keeping $\mathbf{o}$ outside the enlarged span. The seed detectors are checked for the same independence afterward. When a discovered parity fails this test, that direction is the logical observable itself. So, we skip it, and the result is a clean split
\begin{equation}
  \mathcal P \;=\; \mathcal D \,\oplus\, \langle\mathbf o\rangle,
  \label{eq:obs-complement}
\end{equation}
with the detectors on one side and the logical on the other. For logicals $\ell$, the same
test is applied to every $\mathbf o_j$ and every XOR of them, keeping
$\mathbf o_1,\dots,\mathbf o_\ell$ jointly independent of $\mathcal D$ so that no
two logicals silently merge.

\paragraph{From a correct span to a decodable basis.}
Equation~\eqref{eq:parity-space} fixes the detector \emph{space}, but not the right basis. Discovery gives the correct space: dimension, determinism of each detector, gauge count, the complement in Eq.~\eqref{eq:obs-complement}, but it does not control which basis of $\mathcal P$ it returns. On the seam, Discovery returns a cumulative basis. To see what this means, write $\Delta_r$ for the local seam detector at round $r$. A cumulative basis is built from running sums,
\begin{equation}
  D'_r \;=\; \bigoplus_{r'\le r}\Delta_{r'} ,
\end{equation} 
so a single seam fault at round $r$, which should flip only $\Delta_r$, instead flips $D'_r, D'_{r+1}, \dots$ from that round onward. The basis Discovery actually returns is not literally these running sums, but it behaves the same way: before the change of basis below, a single fault can light as many as $18$ detectors on our $d{=}5$ seam. One fault lights a whole tail.

To see why this is a problem, lets model a fault $\varepsilon$ as an indicator vector
$\mathbf f_\varepsilon\in\mathbb F_2^{N}$ over the records it flips. The fault
flips detector $\mathbf d$ when they share an odd number of records,
\begin{equation}
  H_{\varepsilon,\mathbf d} \;=\; \mathbf f_\varepsilon^{\!\top}\mathbf d.
  \label{eq:incidence}
\end{equation}
So, a row of $H$ lists the detectors that a fault lights, which are called its \emph{syndrome}. The minimum-weight perfect matching decoder~\cite{Dennis_2002,Higgott_2025} reads this syndrome on a graph whose nodes are detectors and whose edges are faults. An edge joins two nodes, so every fault must light at most two detectors, or split into pieces that do. A fault that lights a whole tail is a hyperedge, and no such splitting exists. The cumulative basis is thus correct but undecodable by MWPM.

The repair is a change of basis, $\mathbf d_i \leftarrow \mathbf d_i \oplus
\mathbf d_j$. This is an elementary $\mathrm{GL}(\mathbb F_2)$ operation, so the
span $\mathcal D$, its properties, and true code distance are not modified. The change of basis only alters the broken decoding graph, and it recombined detectors without ever removing one.

By the linearity of~\eqref{eq:incidence}, recombining detectors XORs the columns
of $H$ the same way, $\mathrm{col}_{\mathbf d_i} \leftarrow \mathrm{col}_{\mathbf
d_i} \oplus \mathrm{col}_{\mathbf d_j}$. The column of a detector lists the faults
that light it; the row of a fault lists the detectors it lights, and we want each
row to have weight at most two. Both views count the same ones of $H$,
\begin{equation}
  \sum_{\mathbf d}\operatorname{wt}(\mathrm{col}_{\mathbf d})
  \;=\;\sum_{\varepsilon}\operatorname{wt}(\mathrm{row}_{\varepsilon}),
  \label{eq:weight-identity}
\end{equation}
so making the columns sparse makes the rows sparse too. We do this greedily: we
XOR one detector into another whenever it lowers the number of faults that light
it, pushing every fault toward lighting at most two detectors.

The objective is the number of faults per detector, not the number of records per
detector. These differ, and the record count is the wrong target: shrinking a time
edge $\{m_r\oplus m_{r-1}\}$ to a singleton $\{m_r\}$ lowers the record count but
removes a genuine graph edge, which is worse for the decoder.

Finally, a detector that no modeled fault lights up has an all-zero column. It is
fault-redundant, and we drop it in a separate clean-up step. Dropping is not a $\mathrm{GL}(\mathbb F_2)$ operation, so unlike the change of basis it is not guaranteed to preserve $\mathcal D$. It leaves the true distance alone, since the dropped detector caught no modeled fault. But because $\mathcal D$ could in principle change, we re-check the span rank and the complement of Eq.~\eqref{eq:obs-complement} afterward. In our circuits the rank is unchanged and no observable enters the detectors.

\paragraph{Effective distance.}
The graphlike distance reported by \texttt{shortest\_graphlike\_error} is only an upper bound on the true distance. We read and report the true values from \texttt{search\_for\_undetectable\_logical\_errors}. This value is same before and after change of basis. So, the change of basis does not modify the code's parameters. It only makes the code usable for a matching decoder.

\section{Simulation}
\label{sec:simulation}

We simulate the lattice surgery merge for logical Pauli-$XX$ measurement between two distant surface code patches as shown in Figure~\ref{fig:pauliXX-merge}, and the cross-shaped spatial junction merge as shown in Figure~\ref{fig:x-merge}. We run the simulations for distances $d=\{3,5,7,9\}$ for Pauli-XX measurement and $d=\{3,5,7\}$ for the cross shaped spatial junction merge. We designed all the circuits for simulation in Crumble~\cite{crumble}. The circuits were annotated with detectors using the detector annotator that we designed (section~\ref{sec:annotation}) specifically for bell measurement and measurement teleportation seams. The circuit-level detector error model was decoded using the minimum-weight perfect-matching decoder, \texttt{PyMatching}~\cite{Higgott_2025}. We used \texttt{Stim}~\cite{gidney2021stim} to confirm that all detectors are deterministic in the absence of noise, and to run the simulations for calculating the logical error rates. 

We compare the performance of our framework against two instances of the DL interface of Ref.~\cite{Shalby_2025}. In both of these instances, we assign an elevated link error rate $p_{\mathrm{link}}$ to the inter-module two-qubit gates instead of the bulk error rate $p$. In one of the instances, we implement the ``mirror-alternating schedule'' of Ref.~\cite{haug2025latticesurgerybellmeasurements} without its Bell-measurement seam, as in Ref.~\cite{hirai2026}. For the sake of conciseness, we henceforth refer to this ``mirror-alternating schedule'' as just ``alternating schedule,'' and the plots in Section~\ref{sec:results} also use the term ``alternating'' to mean this schedule. We emphasize that it should not be confused with the temporally or spatially alternating schedules mentioned in Section~\ref{sec:prior}. For the second instance, we use the interleaving schedule introduced in Ref.~\cite{hirai2026} for logical Pauli-XX measurement lattice surgery. We also extend their implementation to cross shaped spatial junction merge for comparison.

To compare the performance of the different frameworks, we employ a circuit-level uniform depolarizing noise model. Let $\mathcal{P}_n = \{I, X, Y, Z\}^{\otimes n}$ denote the set of $n$-qubit Pauli operators. We write a general element of $\mathcal{P}_n$ as $P = \sigma_{a_1} \otimes \cdots \otimes \sigma_{a_n}$ with $a_k \in \{I, X, Y, Z\}$, and we denote the $n$-qubit identity operator by $I^{\otimes n}$. After every $n$-qubit gate we apply an $n$-qubit uniform depolarizing
channel to the participating qubits, where each of the $4^n -1$ non-identity Pauli operators occurs with equal probability $p$. Thus, the $n$-qubit uniform depolarizing noise channel acts as: 
\begin{equation}
\mathcal{E}_n^{(p)}(\rho)
= (1 - p)\, \rho
+ \frac{p}{4^n - 1}
\sum_{\substack{P \in \mathcal{P}_n \\ P \neq I^{\otimes n}}}
P\, \rho\, P^{\dagger}
\end{equation}
where, $\rho$ denotes the qubit state, and $p$ denotes the physical error rate.
The identity term occurs with probability $(1-p)$, so the total probability of the non-trivial Pauli error is $p$. This $p$ is distributed uniformly over the $(4^n -1)$ non-identity Pauli operators in the error channel such that the probability of each non-trivial error occurring is equal. Except state preparation and measurement, after every single-qubit, two-qubit, and three-qubit gates, we apply the corresponding single-qubit, two-qubit, and three qubit uniform depolarizing noise channels on the qubits participating in those gate operations. These error channels act as:
\begin{align}
\mathcal{E}_1^{(p)}(\rho)
&= (1 - p)\, \rho
+ \frac{p}{3} \sum_{\sigma \in \{X, Y, Z\}} \sigma\, \rho\, \sigma, \\[4pt]
\mathcal{E}_2^{(p)}(\rho)
&= (1 - p)\, \rho
+ \frac{p}{15}
\sum_{\substack{(\sigma_i, \sigma_j) \\ \neq (I, I)}}
(\sigma_i \otimes \sigma_j)\, \rho\, (\sigma_i \otimes \sigma_j), \\[4pt]
\mathcal{E}_3^{(p)}(\rho)
&= (1 - p)\, \rho
+ \frac{p}{63}
\sum_{P \in \mathcal{P}_3; P \neq I^{\otimes 3}}
P\, \rho\, P^{\dagger}.
\end{align}
where each sum runs over all non-identity Pauli operators on the respective support. The single-qubit channel $\mathcal{E}_1^{(p)}(\rho)$ follows every single-qubit gate, the two-qubit channel $\mathcal{E}_2^{(p)}(\rho)$ follows every two-qubit gate, and the three-qubit channel $\mathcal{E}_3^{(p)}(\rho)$ follows every three-qubit $\texttt{CXX}$ gate. The qubits that stay idle while other qubits participate in gate operations are idle qubits. Such qubits are subject to the single-qubit uniform depolarizing channel $\mathcal{E}_1^{(p)}(\rho)$. Here, we take the idling error rate to be the same as the gate error rate.

We model the state preparation and measurement errors separately from the uniform depolarizing noise channel. Each reset is followed by a bit-flip ($X$) error after preparation in the $\ket{0}$ state or a phase-flip ($Z$) error after preparation $\ket{+}$ state with probability equal to the bulk error rate $p$. To model a noisy measurement, each outcome is flipped with probability equal to the bulk error rate $p$. Since $\texttt{Stim}$ does not support three qubit gates like \texttt{CXX}, we simulate it as two consecutive \texttt{CX} gates, but for noise simulation, we treat it as a single hardware operation occupying a single time step. So, it is followed by a single three-qubit uniform depolarizing noise channel $\mathcal{E}_3^{(p)}(\rho)$ on its support, and its three qubits are not additionally charged idling noise during that step.

Here, our goal is to compare the resources used and the logical error rates of the different frameworks at an elevated link error rate. So, we apply a fixed bulk error rate $p=10^{-4}$ in the depolarizing channels after each single-qubit gate and two-qubit gate and during idle steps. After the inter-module two-qubit and three-qubit gates, we apply an elevated link error rate $p_{\mathrm{link}}$ in the uniform depolarizing channels, whose value we sweep over $10p$ to $100p$.

We estimate the logical error rate by direct Monte Carlo sampling. For each value of the link error rate, we compile the noisy circuit to a detector error model and sample $N = 10^{8}$ shots with \texttt{Stim}. Each shot is decoded by applying \texttt{PyMatching} to the matching graph derived from the detector error model. A shot is counted as a logical failure when the decoder's predicted logical observable disagrees with the observable recorded in simulation, and we report the logical error rate as the fraction of failing shots that are independently evaluated at each link error rate. The three-qubit channel $\mathcal{E}_3^{(p)}(\rho)$ is implemented as a set of mutually exclusive correlated Pauli errors. When building the detector error model, we approximate these as independent and decompose higher-weight errors into graphlike (matchable) components, as required by the matching decoder.




\begin{figure*}[t]
    \centering
    \begin{subfigure}{0.48\textwidth}
        \includegraphics[width=\textwidth]{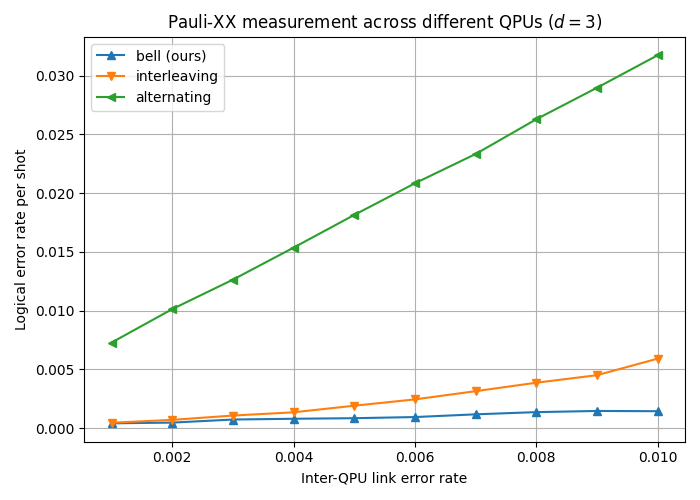}
        \caption{}
        \label{fig:pauliXX_d3}
    \end{subfigure}
    \hfill
    \begin{subfigure}{0.48\textwidth}
        \includegraphics[width=\textwidth]{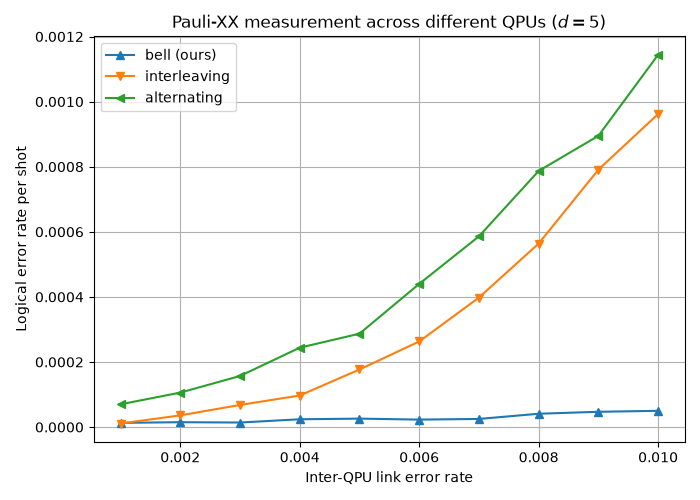}
        \caption{}
        \label{fig:pauliXX_d5}
    \end{subfigure}

    \vskip\baselineskip

    \begin{subfigure}{0.48\textwidth}
        \includegraphics[width=\textwidth]{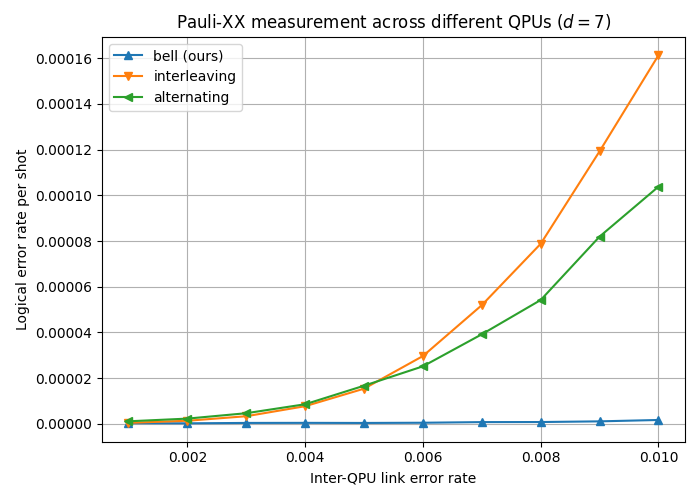}
        \caption{}
        \label{fig:pauliXX_d7}
    \end{subfigure}
    \hfill
    \begin{subfigure}{0.48\textwidth}
        \includegraphics[width=\textwidth]{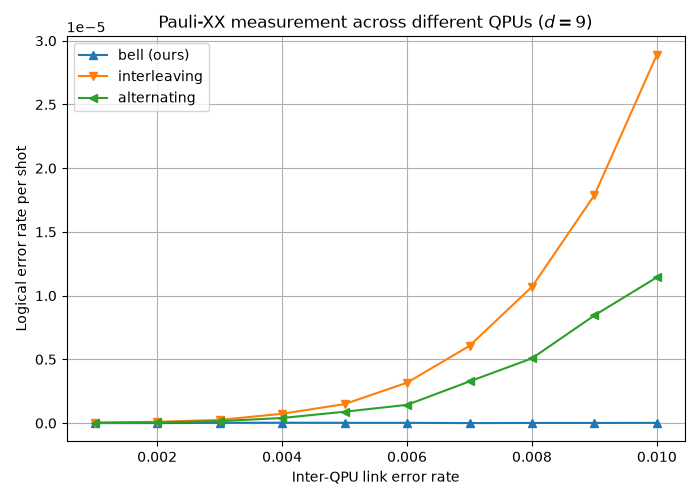}
        \caption{}
        \label{fig:pauliXX_d9}
    \end{subfigure}
    \caption{Logical error rate per shot of the logical Pauli-$XX$ measurement between two surface-code patches on different QPUs, using our Bell-seam framework, the interleaving schedule~\cite{hirai2026}, and the alternating schedule~\cite{haug2025latticesurgerybellmeasurements}, for (a) $d=3$, (b) $d=5$, (c) $d=7$, and (d) $d=9$. The bulk error rate is fixed at $p=10^{-4}$, and the inter-QPU link error rate $p_{\text{link}}$ is swept from $10p$ to $100p$. Each point is estimated from $10^8$ shots.}
    \label{fig:pauliXX_ler}
\end{figure*}

\begin{figure*}[t]
    \centering
    \begin{subfigure}{0.48\textwidth}
        \includegraphics[width=\textwidth]{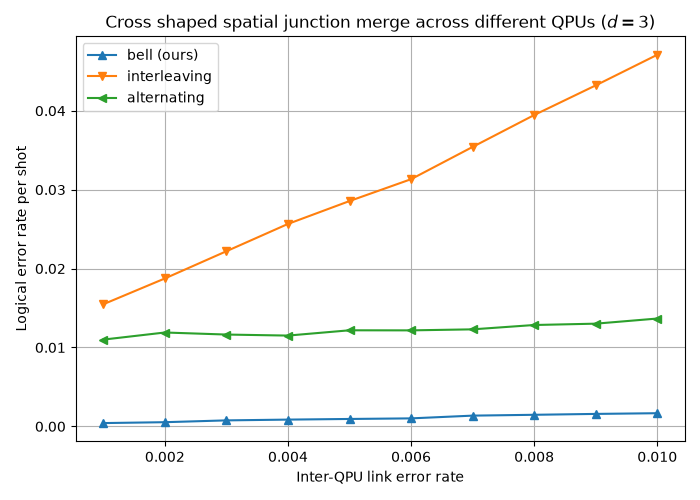}
        \caption{}
        \label{fig:x_merge_d3}
    \end{subfigure}
    \hfill
    \begin{subfigure}{0.48\textwidth}
        \includegraphics[width=\textwidth]{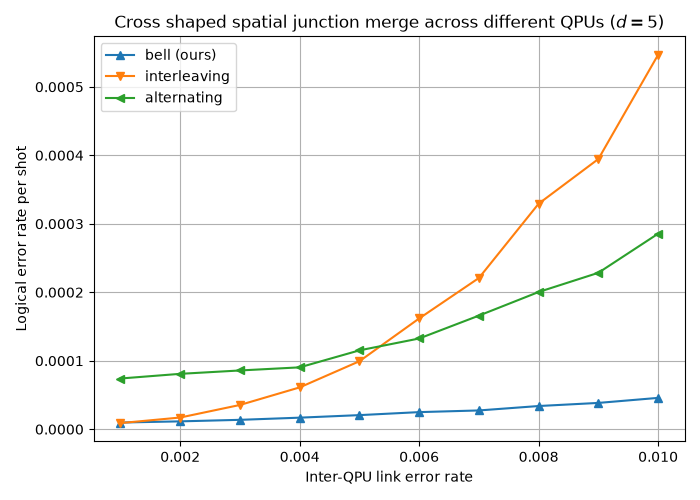}
        \caption{}
        \label{fig:x_merge_d5}
    \end{subfigure}

    \vskip\baselineskip

    \begin{subfigure}{0.48\textwidth}
        \includegraphics[width=\textwidth]{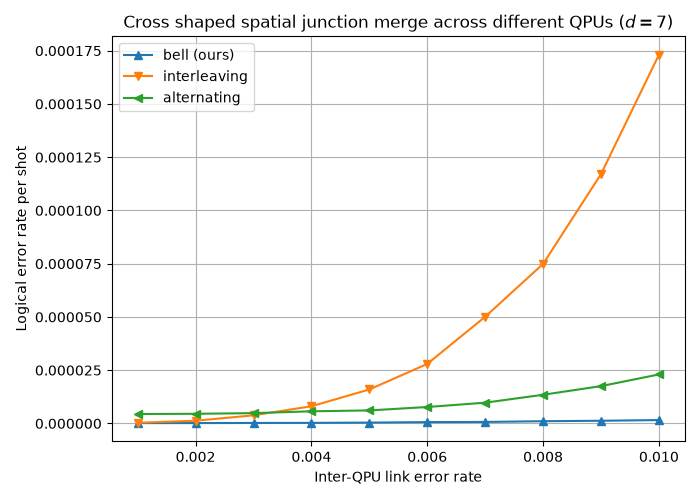}
        \caption{}
        \label{fig:x_merge_d7}
    \end{subfigure}
    \caption{Logical error rate per shot of the cross-shaped spatial junction merge across different QPUs, using our Bell-seam framework, the interleaving schedule~\cite{hirai2026}, and the alternating schedule~\cite{haug2025latticesurgerybellmeasurements}, for (a) $d=3$, (b) $d=5$, and (c) $d=7$. The bulk error rate is fixed at $p=10^{-4}$, and the inter-QPU link error rate $p_{\text{link}}$ is swept from $10p$ to $100p$. Each point is estimated from $10^8$ shots.}
    \label{fig:x_merge_ler}
\end{figure*}

\section{Numerical Results and Discussion}
\label{sec:results}

Here, we present the results obtained from the numerical simulations, and compare our framework with the alternating schedule~\cite{haug2025latticesurgerybellmeasurements} and the interleaving schedule~\cite{hirai2026}, both instantiated as direct-link interfaces (Section~\ref{sec:simulation}) for modular lattice surgery. We compare the logical error rates of these frameworks by sweeping the link error rate at a fixed bulk error rate. We also report the fault distance of our framework and the physical qubit counts of all three frameworks. The Pauli-$XX$ measurement and the cross-shaped junction merge are different logical operations on different layouts, so we compare the frameworks within each layout and not across layouts.

\subsection{Logical error rate of the Pauli-$XX$ measurement lattice surgery:}

Figure~\ref{fig:pauliXX_ler} presents the logical error rate per shot calculated during the logical Pauli-$XX$ measurement lattice surgery for $d = 3, 5, 7, 9$, as $p_{\text{link}}$ is swept from $10p$ to $100p$.

At the lowest link error rate, $p_{\text{link}} = 10p$, the logical error rates of our framework and the interleaving schedule are close at every distance, with ours slightly lower. For example, at $d = 3$ our framework gives $\approx 4.0\times10^{-4}$ and the interleaving schedule gives $\approx 4.5\times10^{-4}$. The alternating schedule is already higher at this point. It gives the logical error rate of $\approx7.3\times10^{-3}$ at $d = 3$ and $\approx 5.5\times10^{-5}$ at $d = 5$, compared to $\approx 4\times10^{-4}$ at $d = 3$ and $\approx 1\times10^{-5}$ at $d = 5$ for our framework.

The performance of the frameworks separates as the link error rate increases. At $d = 3$ (Figure~\ref{fig:pauliXX_d3}), over the full sweep, the logical error rate of the interleaving schedule grows by a factor of about $13$, from $\approx4.5\times10^{-4}$ to $\approx5.9\times10^{-3}$. At the same increase in the link error rate, the logical error rate of our framework grows by a factor of about $3.6$, from $\approx 4\times10^{-4}$ to $\approx 1.45\times10^{-3}$.

At $d=5$ (Figure~\ref{fig:pauliXX_d5}), the logical error rate of our framework stays stably low over the full sweep. Over the same range, the interleaving schedule rises from $\approx1.3\times10^{-5}$ to $\approx9.6\times10^{-4}$, and the alternating schedule from $\approx5.4\times10^{-5}$ to $\approx1.2\times10^{-3}$. At $p_{\text{link}} = 100p$ for $d=5$, our framework gives a logical error rate about $60$ times lower than interleaving schedule and about $75$ times lower than alternating schedule.

The same trend holds at $d=7$ (Figure~\ref{fig:pauliXX_d7}) and $d=9$ (Figure~\ref{fig:pauliXX_d9}). At $d=7$, over the full sweep, the interleaving schedule rises from $\approx4.7\times10^{-7}$ to $\approx1.6\times10^{-4}$, and the alternating schedule from $\approx1.2\times10^{-6}$ to $\approx1.0\times10^{-4}$. Our framework reaches only $\approx1.7\times10^{-6}$ at $p_{\text{link}} = 100p$. At $d=9$, at $p_{\text{link}} = 100p$, the interleaving schedule reaches $\approx2.9\times10^{-5}$ and the alternating schedule $\approx1.2\times10^{-5}$, while our framework stays at $\approx2\times10^{-8}$. At these two distances, the two existing schedules also swap order. The interleaving schedule grows faster with the link error rate and overtakes the alternating schedule, near $p_{\text{link}} = 50p$ at $d=7$ and from about $p_{\text{link}} = 30p$ at $d=9$. Our framework stays below both over the full sweep.

Increasing the code distance still suppresses the logical error rates for all three frameworks, even at $p_{\text{link}} = 100p$. The primary difference that we want to point out is how steeply the logical error rate grows with an increase in the link error rate at a fixed bulk error rate and a fixed code distance for the alternating and interleaving schedules, while it stays low for our framework. The gap also widens with distance. At $p_{\text{link}} = 100p$, our logical error rate is lower than that of the interleaving schedule by a factor of about $4$ at $d=3$, about $60$ at $d=5$, and about $90$ at $d=7$. Hence, our framework gives the lowest logical error rate at every distance and every link error rate we sampled, and it is closest to the interleaving schedule only at $p_{\text{link}} = 10p$.

This behavior of the logical error rates follows from the circuit construction explained in Section~\ref{sec:construction}. In our framework, the Bell-creating and Bell-measuring inter-module CNOTs act only between the two halves of a split ancilla and never touch a data qubit. At the right seam, this CNOT is the last two-qubit gate before the ancillas are measured. So, a link fault on it can only flip the seam measurement outcome, and repeated rounds of syndrome extraction correct it as a measurement error. Additionally, our use of three-qubit gates maintains the CNOT depth at $4$, removing an extra idling step. In the direct-link instances of the alternating and interleaving schedules, the inter-module gates act between an ancilla and a data qubit. So, a link fault there lands on a data qubit, and in the interleaving schedule, the moving tiles carry it further into the bulk (Section~\ref{sec:prior}).

\subsection{Logical error rate of the cross-shaped spatial junction merge:}

Figure~\ref{fig:x_merge_ler} presents the logical error rate per shot of the cross-shaped spatial junction merge for $d = 3, 5, 7$, over the same sweep of $p_{\text{link}}$. Here as well, our framework gives the lowest logical error rate at every distance and every link error rate we sampled.

At $d=3$ (Figure~\ref{fig:x_merge_d3}), both existing schedules already give high logical error rates at $p_{\text{link}} = 10p$: $\approx1.1\times10^{-2}$ for the alternating schedule and $\approx1.5\times10^{-2}$ for the interleaving schedule, compared with $\approx4\times10^{-4}$ for our framework. Over the sweep, the interleaving schedule grows to $\approx4.7\times10^{-2}$. The alternating schedule barely changes and stays between $\approx1.1\times10^{-2}$ and $\approx1.6\times10^{-2}$. So at this distance, its failures come mostly from the bulk rather than the link. Our framework grows to $\approx1.6\times10^{-3}$, which is still about $10$ times lower than the alternating schedule and about $29$ times lower than the interleaving schedule at $p_{\text{link}} = 100p$.

At $d=5$ (Figure~\ref{fig:x_merge_d5}), our framework and the interleaving schedule give nearly the same logical error rate at $p_{\text{link}} = 10p$, $\approx9\times10^{-6}$, while the alternating schedule gives $\approx7.3\times10^{-5}$. Over the sweep, the interleaving schedule grows by a factor of about $60$, to $\approx5.5\times10^{-4}$, and the alternating schedule grows to $\approx2.9\times10^{-4}$. Our framework grows to only $\approx4.5\times10^{-5}$. At $p_{\text{link}} = 100p$, this is about $12$ times lower than the interleaving schedule and about $6$ times lower than the alternating schedule. The alternating and interleaving schedules swap order near $p_{\text{link}} \approx55p$.

At $d=7$ (Figure~\ref{fig:x_merge_d7}), the interleaving schedule rises from $\approx3.2\times10^{-7}$ to $\approx1.7\times10^{-4}$, and the alternating schedule from $\approx4.5\times10^{-6}$ to $\approx2.3\times10^{-5}$. They swap order near $p_{\text{link}} \approx30p$. Our framework stays below both, reaching only $\approx1.6\times10^{-6}$at $p_{\text{link}} = 100p$.


\subsection{Fault distance}

Using \texttt{search\_for\_undetectable\_logical\_errors} (Section~\ref{sec:annotation}), we find that our framework has fault distance $d-1$ at every simulated distance, $d = 3, 5, 7, 9$. This equals the fault distance of the alternating schedule and is one less than the full distance $d$ of the interleaving schedule (Section~\ref{sec:prior}). So the lower logical error rates in Fig.~\ref{fig:pauliXX_ler} do not come from a larger distance. We associated the low logical error rate to the confinement of link faults and careful design to avoid hooks and idling. At equal or smaller distance, fewer of the low-weight failure mechanisms pass through the noisy link.

In our right seam, we implement a modified version of the bell-measurement protocol of Ref.~\cite{haug2025latticesurgerybellmeasurements} by introducing the use of \texttt{CXX} gate. So, we attain a fault distance of $d-1$ with noisy bulk and elevated link error rate while they report a fault distance of only $3d/4$ when the bulk is also noisy. For $d \ge 5$, $d-1$ exceeds $3d/4$.

\subsection{Physical qubit overhead}

Tables~\ref{tab:qubits_pauliXX} and~\ref{tab:qubits_xmerge} list the total number of physical qubits used by each lattice surgery framework compared here. The alternating schedule uses the same number of physical qubits as a standard rotated surface-code patch would use for the given lattice surgery without any additional modification to their structure. So, we use it as a reference to compare the physical qubit overhead of other frameworks. 

\begin{table}[t]
    \centering
    \caption{Total number of physical qubits used by each framework for the logical Pauli-$XX$ measurement lattice surgery. The alternating schedule uses the same qubit count as standard rotated surface-code patches. Our framework uses $2d$ more physical qubits than the alternating schedule, one per Bell pair, and the interleaving schedule uses $2(d+3)$ more. So our framework uses $6$ fewer physical qubits than the interleaving schedule at every distance.}
    \label{tab:qubits_pauliXX}
    \begin{tabular}{c | c | c | c}
        \toprule
        & \multicolumn{3}{c}{Framework} \\
        \cmidrule(lr){2-4}
        $d$ & Bell (ours) & Interleaving & Alternating \\
        \midrule
        3 & 93 & 99 & 87 \\
        5 & 257 & 263 & 247 \\
        7 & 501 & 507 & 487 \\
        9 & 825 & 831 & 807 \\
        \bottomrule
    \end{tabular}
\end{table}

Our framework adds exactly one physical qubit per split ancilla, namely the second half of the split ancilla, and we have $d$ bell pairs on each seam. In the two lattice surgery experiments simulated in this research, there are two seams: one on the right and one on the left. So, for the Pauli-$XX$ measurement lattice surgery, our framework uses $2d$ more physical qubits than the alternating schedule. Relative to the alternating schedule, this overhead falls from $6.9\%$ at $d = 3$ to $2.2\%$ at $d = 9$, because it grows linearly in $d$ while the total number of qubits grows as $d^2$. The interleaving schedule uses $2(d+3)$ more physical qubits than the alternating schedule, which is $6$ more than ours at every distance.

\begin{table}[t]
    \centering
    \caption{Total number of physical qubits used by each framework for the cross-shaped spatial junction merge. Our framework uses $2d$ more physical qubits than the alternating schedule, one per Bell pair. The interleaving schedule uses $9$ more physical qubits than ours at $d = 3, 5$ and $11$ more at $d = 7$.}
\label{tab:qubits_xmerge}
    \begin{tabular}{c | c | c | c}
        \toprule
        & \multicolumn{3}{c}{Framework} \\
        \cmidrule(lr){2-4}
        $d$ & Bell (ours) & Interleaving & Alternating \\
        \midrule
        3 & 117 & 126 & 111 \\
        5 & 297 & 306 & 287 \\
        7 & 557 & 568 & 543 \\
        \bottomrule
    \end{tabular}
\end{table}

The cross-shaped spatial junction merge as well has two seams on either sides. So, our framework uses $2d$ more physical qubits than the alternating schedule. Relative to the alternating schedule, this overhead falls from $5.4\%$ at $d = 3$ to $2.6\%$ at $d = 7$. The interleaving schedule uses $15$, $19$, and $25$ more physical qubits than the alternating schedule at $d = 3, 5, 7$, respectively which is respectively $9$, $9$, and $11$ more than ours.

So in both layouts, our overhead over the standard qubit count is $2d$, which is linear in $d$ and stays below $7\%$ at every distance we simulated, while the interleaving schedule costs more physical qubits than our framework and gives a higher logical error rate at every link error rate we sampled.

\subsection{Summary}

We performed numerical simulations of both lattice surgery layouts at different distances by increasing the link error rate from $10p$ to $100p$. We find that our framework gives the lowest logical error in comparison to the compared frameworks. As $p_{\mathrm{link}}$ increases, the logical error rate of the interleaving and alternating schedule grows sharply, while it stays relatively stably low for our framework. 

Our framework gives this low logical error rate at the fault distance $d-1$, which is one less than the interleaving schedule. So, the advantage we gain against the compared frameworks is a combination of hook protection and confinement of the elevated error of the seam at the seam so that it doesn't spread into the bulk of the code. 

Additionally, our framework keeps the CNOT depth at $4$, uses only $d$ Bell pairs per seam compared with $2d$ Bell pairs used in the in measurement teleportation~\cite{Jacinto_2026}, and uses fewer physical qubits than the interleaving schedule. The cost is a native three-qubit $\texttt{CXX}$ gate at the seam, which has been shown to be feasible with realistic simulated parameters~\cite{tasler2025optimizingsuperconductingthreequbitgates,Old_2025}. Our framework is also simple to design for any complicated structure of the surface code. It can be used to break down any large surface code patch with complicated boundary structure to allow each piece to keep the standard N/Z shaped schedule suited to its own boundaries, and only the seam checks follow a fixed rule. The interleaving schedule instead requires its moving tiles to be steered toward boundaries of the matching Pauli type. In layouts with non-standard boundaries, this needs extra boundary qubits and a careful construction of the boundary to maintain the stabilizer structure. This can be clearly seen in the physical qubit overhead listed for interleaving schedule in Tables~\ref{tab:qubits_pauliXX} and~\ref{tab:qubits_xmerge}. Hence, our framework gives a practical route to lattice surgery across QPUs whose links are up to two orders of magnitude noisier than the bulk.

\section{Conclusion and Outlook}
\label{sec:outlook}

In this paper, we introduce a new lattice surgery framework for modular quantum computation that gives the lowest logical error rate at elevated link error rates, while still maintaining minimum CNOT depth. Using this framework, we can split a large surface code with complicated boundary structures into smaller patches of uniform boundaries, assign hook avoiding N/Z shaped schedules to each of those patches, and join them all together using Bell-measurement and measurement-teleportation at the seam. We use three-qubit $\texttt{CXX}$ gate at the seam for parity measurement to maintain minimum CNOT depth of the circuit and avoid idling. We use $d$ bell pairs per seam and thus additional $d$ physical qubits per seam compared to a standard rotated surface code. We also gave an algebraic detector annotator for such seams, which recovers the full detector space from the circuit's measurement statistics in a basis a matching decoder can use.

We demonstrate our framework on lattice surgery for logical Pauli-$XX$ measurement of distant surface code patches and the cross-shaped spatial junction merge, at a bulk error rate $p = 10^{-4}$, sweeping the link error rate from $10p$ to $100p$. In both layouts, our framework gives the lowest logical error rate in comparison to the other frameworks compared here, at every distance. Our framework gives a fault distance $d-1$, works at CNOT depth $4$, uses $d$ Bell pairs per seam, and $2d$ physical qubits more than a standard rotated surface code.

We compared the performance of lattice surgery frameworks using a uniform depolarizing noise model. We reported that our framework gives a fault distance of $d-1$. This disadvantage could be compensated if this framework is used in lattice surgery with a dephasing noise model. Studying the performance of our framework in such biased noise scenario can be pursued as a future research.

Another clear future direction of this work is to increase the fault distance from $d-1$ to $d$. Searching for undetectable logical errors, we found error the possible chains of length $d-1$ that result in this fault distance. This suggests that a single error is being copied onto two data qubits to reduce the fault distance. The specific fault location can be pin pointed and flag qubits can be added at these locations to ensure the fault distance is the full code distance. This could further suppress the logical error rates.

Furthermore, creating a fast detector annotator that can accurately annotate the non-trivial lattice surgery structures that use Bell pairs at the seam is an open problem. In this research, we algebraically annotate our construction, and was not substantially expensive for the distances simulated here. But our detector annotation method could be computationally expensive for larger distances. So, creating a fast detector annotator for such seam constructions is clearly a non trivial task, and creating a proper algorithm to do it fast can be pursued as a future research.

\section{Acknowledgement}
This material is based upon work supported by Argonne as part of the LDRD project Distributed Quantum Computing Algorithms Lab.

\section{Code Availability}

The Stim circuits and the codes used for the simulations in this research can be found in this github repository:
\href{https://github.com/Pritesh402/modular-surgery}{https://github.com/Pritesh402/modular-surgery}

\vfill

\small

\noindent\framebox{\parbox{0.97\linewidth}{
The submitted manuscript has been created by UChicago Argonne, LLC, Operator of 
Argonne National Laboratory (``Argonne''). Argonne, a U.S.\ Department of 
Energy Office of Science laboratory, is operated under Contract No.\ 
DE-AC02-06CH11357. 
The U.S.\ Government retains for itself, and others acting on its behalf, a 
paid-up nonexclusive, irrevocable worldwide license in said article to 
reproduce, prepare derivative works, distribute copies to the public, and 
perform publicly and display publicly, by or on behalf of the Government.  The 
Department of Energy will provide public access to these results of federally 
sponsored research in accordance with the DOE Public Access Plan. 
http://energy.gov/downloads/doe-public-access-plan.}}

\bibliographystyle{quantum}
\bibliography{mybibliography}

\end{document}